\documentstyle[12pt]{article}
\input epsf
\begin{document}
\def\hw {\hbar \omega}
\def\ie{{\em i.e., }}
\def\rf#1{{(\ref{#1})}}
\def\mbs{\mbox{\boldmath$\sigma$}}
\def\gsim{\:\raisebox{-0.5ex}{$\stackrel{\textstyle>}{\sim}$}\:}
\def\lsim{\:\raisebox{-0.5ex}{$\stackrel{\textstyle<}{\sim}$}\:}
\def\Ket#1{||#1 \rangle}
\def\Bra#1{\langle #1||}
\def\ubp{\bar{u}_{{\rm p}}}
\def\vbp{\bar{v}_{{\rm p}}}
\def\ubn{\bar{u}_{{\rm n}}}
\def\vbn{\bar{v}_{{\rm n}}}
\def\up{u_{{\rm p}}}
\def\vp{v_{{\rm p}}}
\def\un{u_{{\rm n}}}
\def\vn{v_{{\rm n}}}
\def\rop{\rho_{{\rm p}}}
\def\ron{\rho_{{\rm n}}}
\def\mbn{\mbox{\boldmath$\nabla$}}
\def\sss{\scriptscriptstyle}
\def\gp{g_{\sss +}}
\def\gm{g_{\sss -}}
\def\endauthors{}
\def\authors#1\endauthors{#1}
\begin{titlepage}
\pagestyle{empty}
\baselineskip=21pt
\vskip .2in
\begin{center}
{\large{\bf Violation of the Ikeda Sum Rule and the Self-consistency in the
Renormalized Quasiparticle Random Phase Approximation and the Nuclear
Double-beta Decay}}$^*$
 \end{center}
\vskip .1in

\authors
\centerline{ F. Krmpoti\'{c}$^{a\dagger}$, T.T.S. Kuo${}^b$,
A. Mariano$^{a\dagger}$, E.J.V. de Passos${}^c {}^{\dagger\dagger}$ and
A.F.R. de Toledo Piza${}^c$}
\vskip .15in
\centerline{\it ${}^a$ Departamento de F\'\i sica, Facultad de Ciencias
Exactas}
\centerline{\it Universidad Nacional de La Plata, C. C. 67, 1900 La Plata,
Argentina.}
\vskip .1in
\centerline{\it ${}^b$ Physics Department, State University of New
York at Stony Brook}
\centerline {\it Stony Brook, NY 11794-3800, USA.}
\vskip .15in
\centerline{\it ${}^c$  Instituto de F\'\i sica, Universidade de
S\~{a}o Paulo,}
\centerline{\it C.P. 66318, 05389-970 S\~{a}o Paulo, Brasil.}
\endauthors

\indent\indent PACS numbers: 23.40.Hc, 21.10.Re, 21.60.Jz

\centerline{ {\bf Abstract} }

The effect of the inclusion of ground state correlations into the QRPA
equation of motion for the two-neutrino double beta
($\beta\beta_{2\nu}$) decay is carefully analyzed.  The resulting
model, called renormalized QRPA (RQRPA), does not collapse near the
physical value of the nuclear force strength in the particle-particle
channel, as happens with the ordinary QRPA. Still, the
$\beta\beta_{2\nu}$ transition amplitude is only slightly less
sensitive on this parameter in the RQRPA than that in the plain QRPA.
It is argued that this fact reveals once more that the characteristic
behaviour of the $\beta\beta_{2\nu}$ transition amplitude within the
QRPA is not an artifact of the model, but a consequence of the partial
restoration of the spin-isospin $SU(4)$ symmetry. It is shown that the
price paid for bypassing the collapse in the RQRPA is the violation of
the Ikeda sum rule.
\baselineskip=18pt
\bigskip

\noindent
$^*$Work partially supported by the Fundaci\'on Antorchas, Argentina,
and the CONICET from Argentina.
\\ $^{\dagger}$Fellow of the CONICET from Argentina.
\\ ${}^{\dagger\dagger}$Fellow of the CNPq from Brazil.
\end{titlepage}
\baselineskip=18pt

\section {Introduction}
\label{sec1}
The quasiparticle random phase approximation (QRPA) is the most
frequently used nuclear structure method for evaluating $\beta\beta$
rates both for the two-neutrino decay mode ($\beta\beta_{2\nu}$) and
for the neutrinoless mode ($\beta\beta_{0\nu}$). The general feature
of this method is that the resulting nuclear matrix elements ${\cal
M}_{2\nu}$ and ${\cal M}_{0\nu}$ turn out to be highly sensitive to
the particle-particle force in the $S=1$, $T=0$ channel
\cite{Vog86,Civ87,Eng88,Mut88,Hir90a,Krm93}.
Furthermore, the QRPA collapses very close to the
physical value for this force.  One thus may suspect that this method
yields relatively small values of ${\cal M}_{2\nu}$ simply because the
approximation breaks up. In other words, the smallness of ${\cal
M}_{2\nu}$ in the QRPA could be just an artifact of the model.

Several modifications of the QRPA have been proposed in order to amend
the above behavior in a qualitative way, including higher order RPA
corrections \cite{Rad91}, nuclear deformation \cite{Rad93},
single-particle self-energy BCS terms \cite{Kuo92}, particle number
projection \cite{Krm93a} and the proton-neutron pairing \cite{Che93}.
Yet, none of these inhibits the collapse, which is just the famous
"phase transition" where the RPA develops zero or imaginary frequency
solutions.  One should remember in this connection that in the
derivation of the ordinary RPA it is assumed that we can replace, when
evaluating the equations of motion, the RPA (correlated) ground state
by the Hartree-Fock ground state.  Discrepancies due to this
replacement will obviously get more serious the more significant are
the ground state correlations. Attempts to correct this have been made
by embodying the effect of GSC in the RPA equations of motion
\cite{Row68,Sch71,Kar93,Cat94}. The corresponding formalism, named
self-consistent or renormalized RPA (RRPA), generally yields better
results in the sense that the instability develops at a larger
interaction strength or is fully taken away.

Recently, the renormalized QRPA (RQRPA) has been applied to the
$\beta\beta_{2\nu}$ decays by Toivanen and Suhonen \cite{Toi95} (in
$^{100}Mo$) and by Schwieger, \v Simkovic and Faessler \cite{Sch96}
(in $^{76}Ge$, $^{82}Se$, $^{128}Te$ and $^{130}Te$ nuclei). We have
also discussed the $^{100}Mo$ $2\nu$-decay but only in the framework
of a schematic model \cite{Krm96}.  In this work we present a detailed
study of several $\beta\beta$ decaying nuclei within the RQRPA. Particular
attention is given to the non-conservation of the non energy weighted
sum rule (called Ikeda sum rule for $J^{\pi}=1^+$) within this
approximation. The self-consistency between the residual interaction
and the mean field \cite{Hir90a} is also addressed.

\section {Formalism}
\label{sec2}

The RQRPA formalism has been presented in several articles
\cite{Cat94,Toi95,Sch96,Krm96} and we will therefore only sketch the
main formulae needed in the discussion.

\subsection {RQRPA Equation and the Transition Moments}
\label{sec2.1}

The excited states $|\lambda JM\rangle$ are built by the action of the
charge-exchange operators

\begin{equation}
\Omega^{\dag}(\lambda JM) = \sum_{pn} \left[ X_{pn}(\lambda J)
 A^{\dag}_{pn}(JM) - Y_{pn}(\lambda J) A_{pn}(\overline{JM})\right],
\label{1}\end{equation}

\noindent on the QRPA correlated ground state $|\tilde{0}\rangle$.
Here $A^{\dag}_{pn}(JM) =[\alpha^{\dagger}_p\alpha^{\dagger}_n]^{JM}$,
$A_{pn}(\overline{JM})=(-1)^{J+M}A_{pn}(J-M)$, and
$\alpha^{\dagger}_t$ are quasiparticle creation operators for protons
($t=p$) and neutrons ($t=n$).  The amplitudes $X$ and $Y$ and the
eigenvalues $\omega_\lambda$ satisfy the RQRPA equations

\begin{eqnarray}
\left(\begin{array}{ll} {\sf A}(J) & {\sf B}(J) \\ {\sf B}^{*}(J) &
{\sf A}^{*}(J)\end{array}\right) \left(\begin{array}{l} {\sf
X}(\lambda J) \\ {\sf Y}(\lambda J) \end{array}\right) =
\omega_{\lambda J} \left(\begin{array}{l} ~{\sf X}(\lambda J) \\ -
{\sf Y}(\lambda J) \end{array}\right),
\label{2} \end{eqnarray}

\noindent where

\begin{equation}
{\sf X}_{pn}(\lambda J)\equiv X_{pn}(\lambda J) D_{pn}^{1/2}~~~
\mbox{and}~~~{\sf Y}_{pn}(\lambda J)\equiv Y_{pn}(\lambda J)
D_{pn}^{1/2}
\label{3}
\end{equation}

\noindent are the renormalized amplitudes,

\begin{equation}
D_{pn}=1- {\cal N}_p -{\cal N}_n
\label{4}
\end{equation}

\noindent and

\begin{eqnarray}
{\cal N}_t &=& \hat{j}_t^{-1} \langle
\tilde{0}|[\alpha^\dagger_t\alpha_{\bar{t}}]^0 |\tilde{0}\rangle
\label{5}
\end{eqnarray}

\noindent are the quasiparticle occupations
($\hat{J}\equiv\sqrt{2J+1}$). Here we are assuming that the single
quasi-particle density is diagonal, $\langle\tilde{0}|
[\alpha^\dagger_t\alpha_{ \bar{t'}}]^0|\tilde{0}\rangle=
\delta_{tt'}\langle\tilde{0}|[\alpha^\dagger_t
\alpha_{\bar{t}}]^0|\tilde{0}\rangle$. The submatrices ${\sf A}(J)$
and ${\sf B}(J)$ are

\begin{eqnarray}
{\sf A}_{pn,p'n'}(J) & = &
 (\epsilon_p+\epsilon_n)\delta_{pp'}\delta_{nn'} +
 D_{pn}^{1/2}\left[F(pn,p'n',J) (u_pv_nu_{p'}v_{n'}+v_pu_nv_{p'}u_{n}
 ) \right.\nonumber \\ &+ &\left. G(pn,p'n',J)
 (u_pu_nu_{p'}u_{n'}+v_pv_nv_{p'}v_{n'})\right] D_{p'n'}^{1/2},
 \nonumber \\ {\sf B}_{pn,p'n'}(J)& = & D_{pn}^{1/2}\left[F(pn,p'n',J)
 (v_pu_nu_{p'}v_{n'}+u_pv_nv_{p'}u_{n'}) \right. \nonumber \\ & -
 &\left. G(pn,p'n',J)(u_pu_nv_{p'}v_{n'} + v_pv_nu_{p'}u_{n'})\right]
 D_{p'n'}^{1/2},
\label{6} \end{eqnarray}

\noindent where $F$ and $G$ are the usual particle-hole ({\it ph}) and
particle-particle ({\it pp}) coupled two-body matrix elements.

The QRPA ground state is determined by the condition 

\begin{equation}
\label{7}
\Omega(\lambda JM)|\tilde{0}\rangle=0.
\end{equation}

\noindent Writing the ground state in the form 

\begin{equation}
|\tilde{0}\rangle = N_0 e^{{\cal S}} |BCS\rangle,
\label{8} \end{equation}

\noindent using equation (\ref{7}) and making the quasi-boson
approximation, we find

\begin{equation}
{\cal S} = \frac{1}{2} \sum_{pnp'n'J} \hat{J}^{-1}
D_{pn}^{-1/2}
{\sf C}_{pnp'n'}(J)
D_{p'n'}^{-1/2}\left[A^{\dag}_{pn}(J)A^{\dag}_{p'n'}(J)\right]^0 ,
\label{9} \end{equation}

\noindent where the matrix ${\sf C}$ is the solution of 

\begin{equation}
\sum_{pn} {\sf X}_{pn}^*(\lambda J){\sf C}_{pnp'n'}(J)
= {\sf Y}_{p'n'}^*(\lambda J), ~\mbox{for all $\lambda J$}.
\label{10}
\end{equation}

\noindent From these equations one gets that
\begin{eqnarray}
{\cal N}_p &=&\sum_{\lambda J n'}\hat{J}^2\hat{j}_p^{-2}D_{pn'}|{\sf
Y}_{pn'} (\lambda J)|^2, \nonumber\\ {\cal N}_n&=& \sum_{\lambda J
p'}\hat{J}^2\hat{j}_n^{-2}D_{p'n}|{\sf Y}_{p'n} (\lambda J)|^2.
\label{11}
\end{eqnarray}

To evaluate the transition matrix elements

\begin{equation}
\langle\lambda J||{\cal O}_{\sss \pm}(J)||\tilde{0}\rangle =
\langle \tilde{0}|\left[\Omega(\lambda \bar{J}),
{\cal O}_{\sss \pm}(J)\right]^0| \tilde{0}\rangle,
\label{12}
\end{equation}

\noindent the $\beta^{\sss \pm}$ decay operators

\begin{equation}
{\cal O}_{\sss \pm}(JM)=\sum_iO(JM;i)t_{\sss \pm}(i);~~~~
(\mbox{with}~ t_{-}|n\rangle= |p\rangle)
\label{13}
\end{equation}

\noindent are expressed in the form

\begin{equation}
{\cal O}_{\sss \pm}(JM)= {\cal O}^{20}_{\sss \pm}(JM)
+ {\cal O}^{11}_{\sss \pm}(JM),
\label{14}
\end{equation}

\noindent where

\begin{eqnarray}
{\cal O}^{20}_{\sss \pm}(JM)&=&\sum_{pn}\left[\Lambda^{\sss
\pm}_{pn}(J) A^{\dag}_{pn}(JM) +(-1)^{J} \Lambda^{{\sss
\mp}*}_{pn}(J)A_{pn}(\overline{JM})\right],
\nonumber \\
{\cal O}^{11}_{\sss \pm}(JM)&=& \sum_{pn}\left[\Delta^{\sss \pm}_{pn} (J)
B^{\dag}_{pn}(JM) +(-1)^{J} \Delta^{{\sss
\mp}*}_{pn}(J)B_{pn}(\overline{JM})\right],
\label{15}
\end{eqnarray}

\noindent with

\begin{eqnarray}
\Lambda^{\sss +}_{pn}(J)&=&-(-1)^{J}\hat{J}^{-1}v_pu_n \langle
p||O(J)||n\rangle^*,~~~ \Lambda^{\sss -}_{pn}(J)=-\hat{J}^{-1}u_pv_n
\langle p||O(J)||n\rangle, \nonumber\\ \Delta^{\sss
+}_{pn}(J)&=&-(-1)^{J}\hat{J}^{-1}v_pv_n \langle
p||O(J)||n\rangle^*,~~~ \Delta^{\sss
-}_{pn}(J)=\hat{J}^{-1}u_pu_n\langle p||O(J)||n\rangle,
\label{16}
\end{eqnarray}

\noindent $B^{\dag}_{pn}(JM)=[\alpha^{\dagger}_p 
\alpha_{\bar{n}}]^{JM}$ and
$B_{pn}(\overline{JM})=(-1)^{J+M}B_{pn}(J-M)$. Notice that, whereas
the operator ${\cal O}^{20}_{\sss \pm}(JM)$ creates and destroys a
$np$ quasi-particle pair, the scattering operator ${\cal O}^{11}_{\sss
\pm}(JM)$ conserves the number of quasi-particles.

The evaluation of (\ref{12}) goes now via the following relationships:

\begin{eqnarray}
\hat{J}^{-1} \langle \tilde{0}|
\left[A_{pn}(\bar{J}),A^{\dag}_{p'n'}(J)\right]^0 |\tilde{0}\rangle &
=& \delta_{pp'}\delta_{nn'} D_{pn}, \nonumber\\ \hat{J}^{-1}\langle
\tilde{0}| \left[B_{pn}(\bar{J}),B^{\dag}_{p'n'}(J)\right]^0|\tilde{0}
\rangle &=& \delta_{pp'}\delta_{nn'} ({\cal N}_n -{\cal N}_p),
\nonumber\\ \hat{J}^{-1}\langle\tilde{0}|\left[A^{\dag}_{pn}(J),
B^{\dag}_{p'n'}(J)\right]^0 | \tilde{0} \rangle & = &
-\delta_{nn'}\hat{j}_p^{-1} \langle \tilde{0} | \left[\alpha^\dagger_p
\alpha^\dagger_{p'}\right]^0| \tilde{0} \rangle\equiv 0, \nonumber\\
\hat{J}^{-1}\langle\tilde{0}|\left[
A^{\dag}_{pn}(J),B_{p'n'}(\bar{J})\right]^0| \tilde{0} \rangle & = &-
\delta_{pp'}\hat{j}_n^{-1} \langle \tilde{0}|
\left[\alpha^\dagger_n\alpha^\dagger_{n'}\right]^0 |\tilde{0}
\rangle\equiv 0.
\label{17} 
\end{eqnarray}

\noindent The expectation values
$\langle\tilde{0}|\left[\alpha^\dagger_t \alpha^\dagger_{t'}\right]^0
|\tilde{0} \rangle$ are identically zero since $|\tilde{0} \rangle$ is
a superposition of states with equal number of neutron and proton
quasi-particles (see eqs. (\ref{8}) and (\ref{9})).   Thus, only the
first term in eq. (\ref{14}) contributes to the transition matrix
element and one gets

\begin{eqnarray}
\langle\lambda J||{\cal O}_{\sss \pm}(J)||\tilde{0}\rangle &\equiv&
\langle \tilde{0}|\left[\Omega(\lambda \bar{J}), {\cal O}^{20}_{\sss
\pm}(J)\right]^0| \tilde{0}\rangle\nonumber\\ &=&\hat{J}
\sum_{pn}\left[\Lambda^{\sss \pm}_{pn}(J) {\sf X}^*_{pn}(\lambda J) +
(-1)^J \Lambda^{{\sss \pm}*}_{pn}(J){\sf Y}^*_{pn}(\lambda
J)\right]D_{pn}^{1/2}
\label{18}
\end{eqnarray}

\noindent The corresponding RQRPA total strengths are

\begin{equation}
\tilde{S}_{\sss \pm}(J)=\hat{J}^{-2} \sum_{\lambda}|\langle\lambda
J||{\cal O}_{\sss \pm}(J)||\tilde{0}\rangle|^2,
\label{19}
\end{equation}

\noindent and from (\ref{18}) we obtain that

\begin{eqnarray}
\tilde{S}_{\sss -}(J)-\tilde{S}_{\sss +}(J)
&=&\sum_{pn}(|\Lambda^{\sss -}_{pn}(J)|^2 -|\Lambda^{\sss
+}_{pn}(J)|^2)D_{pn}\nonumber\\ &=&\hat{J}^{-2}
\sum_{pn}(v_n^2-v_p^2)D_{pn} |\langle p||O(J)||n\rangle|^2.
\label{20}
\end{eqnarray}

For the evaluation of the $\beta\beta_{2\nu}$ matrix elements, instead
of the two-vacua QRPA method introduced in ref. \cite{Hir90}, we will
here simply use the expression

\begin{equation}
{\cal M}_{2\nu}(J)=\sum_{\lambda}\frac{\langle\tilde{0}||{\cal
 O}_{\sss -}(J)|| \lambda J\rangle \langle\lambda J||{\cal O}_{\sss
 -}(J)||\tilde{0}\rangle} {\omega_{\lambda J}},
\label{21}
\end{equation}

\noindent but solving the gap equations for the intermediate nucleus
\cite{Cha83}.  We have tested numerically that both methods yield
almost identical results.

\subsection {Sum Rule}
\label{sec2.2}
It is well known that the total $\beta^{\pm}$ strengths $S_{\sss \pm}(J)$

\begin{equation}
S_{\sss \pm}(J)=\hat{J}^{-2}
\sum_{\nu}|\langle\nu J||{\cal O}_{\sss \pm}(J)||0\rangle|^2
\label{22}
\end{equation}

\noindent can be expressed in the form

\begin{equation}
S_{\sss \pm}(J)=(-)^J\hat{J}^{-1}\langle 0|[{\cal O}_{\sss
\mp}(J){\cal O}_{\sss \pm}(J)]^0|0\rangle,
\label{23}
\end{equation}

\noindent when $|\nu J\rangle$ is the complete set of excited states
that can be reached by operating with ${\cal O}_{\sss \mp}(J)$ on the initial
state $|0\rangle$. It follows at once that

\begin{equation}
S_{\sss -}(J) -S_{\sss +}(J)= (-)^{J}\hat{J}^{-1} \langle 0|[{\cal
O}_+(J),{\cal O}_-(J)]^0| 0\rangle,
\label{24}
\end{equation}

\noindent \ie the difference between the exact total $S_-(J)$ and
$S_+(J)$ strengths equals the expectation value of the operator
$(-)^{J}\hat{J}^{-1}[{\cal O}_+(J),{\cal O}_-(J)]^0$ in the
ground state $| 0\rangle$.
In particular, for the observables of interest here,
\ie the Fermi (F) and Gamow-Teller (GT) operators $t_{\sss \pm}$ and
$t_{\sss \pm}\sigma$, one gets \cite{Boh75,Krm83}

\begin{equation}
S_-(J^{\pi}=0^+,1^+)-S_+(J^{\pi}=0^+,1^+)=\langle 0| \hat N-\hat Z
|0\rangle.
\label{25} 
\end{equation}

\noindent Moreover, if the ground state is chosen so that

\begin{equation}
\langle 0| \hat N-\hat Z |0\rangle=N-Z,
\label{26} \end{equation}

\noindent one obtains the well known result

\begin{equation}
S_{\sss -}(J^{\pi}=0^+,1^+) -S_{\sss +}(J^{\pi}=0^+,1^+)=N-Z.
\label{27} \end{equation}

\noindent It can now be easily shown that the relations (\ref{23}) and
(\ref{24}) are not valid within the RQRPA, and this leads to the violation
of the Ikeda sum rule.
In fact, using eqs. (\ref{17}) the right hand side of eq. (\ref{24})
evaluates in the RQRPA to 

\begin{eqnarray}
(-)^J\hat{J}^{-1}\langle\tilde{0}|[{\cal O}_+(J), {\cal
O}_-(J)]^0|\tilde{0}\rangle &=&(-)^{J}\hat{J}^{-1}
\langle \tilde{0}|[{\cal O}^{20}_{\sss +}(J),
{\cal O}^{20}_{\sss -}(J)]^0|\tilde{0}\rangle \nonumber \\
&+&(-)^{J}\hat{J}^{-1}
\langle \tilde{0}|[{\cal O}^{11}_{\sss +}(J),
{\cal O}^{11}_{\sss -}(J)]^0|\tilde{0}\rangle.
\label{28} 
\end{eqnarray}

\noindent Calculating the first term on the right hand side of eq. (\ref{28})
one finds that it reproduces the difference between the total RQRPA strengths
$\tilde{S}_-(J)$ and $\tilde{S}_+(J)$ given in eq. (\ref{20}), that is

\begin{equation}
(-)^{J}\hat{J}^{-1} \langle \tilde{0}|[{\cal O}^{20}_{\sss +}(J),
{\cal O}^{20}_{\sss -}(J)]^0|\tilde{0}\rangle = \hat{J}^{-2}
\sum_{pn}(v_n^2-v_p^2)D_{pn} |\langle p||O(J)||n\rangle|^2 =
\tilde{S}_-(J)-\tilde{S}_+(J).
\label{29}
\end{equation}

\noindent As for the second term on the right hand side of
eq. (\ref{28}) one obtains

\begin{equation}
(-)^{J}\hat{J}^{-1} \langle \tilde{0}|[{\cal O}^{11}_{\sss +}(J),
{\cal O}^{11}_{\sss -}(J)]^0|\tilde{0}\rangle =\hat{J}^{-2} \sum_{pn}
|\langle p||O(J)||n\rangle|^2 (u_n^2 - v_p^2)({\cal N}_n -{\cal N}_p).
\label{30}
\end{equation}

\noindent This quantity is different from zero because the states
${\cal O}^{11}_{\sss \pm}(JM)|\tilde{0}\rangle$
are both non-null and orthogonal to the RQRPA model space (spanned by
the states $\Omega^{\dag}(\lambda JM) |\tilde{0}\rangle$).

Using the above results we can express
$\tilde{S}_-(J)-\tilde{S}_+(J)$ for F and GT transitions as

\begin{eqnarray}
\tilde{S}_-(J^{\pi}=0^+,1^+)&-&\tilde{S}_+(J^{\pi}=0^+,1^+)= \langle
\tilde{0}| \hat N-\hat Z|\tilde{0}\rangle \nonumber \\ &-&\hat{J}^{-2}
\sum_{pn} |\langle p||O(J^{\pi}=0^+,1^+)||n\rangle|^2 (u_n^2 -
v_p^2)({\cal N}_n -{\cal N}_p),
\label{31}
\end{eqnarray}

\noindent and the violation of the corresponding sum rules within the RQRPA is
associated with a nonvanishing value of the quantity

\begin{equation}
\Delta\tilde{S} (J^{\pi}=0^+,1^+)=(N-Z) -\left[\tilde{S}_-
(J^{\pi}=0^+,1^+)- \tilde{S}_+ (J^{\pi}=0^+,1^+)\right].
\label{32}
\end{equation}

\noindent As discussed in Sect. 3 below, this depends quantitatively
on the way the gap equations are solved. In fact, the usual
constraints \cite{Toi95,Sch96}

\begin{eqnarray}
N=\langle BCS|\hat{N}|BCS\rangle&=& \sum_n\hat{j}_n^2v_n^2,
\nonumber\\ Z=\langle BCS|\hat{Z}|BCS\rangle&=&
\sum_p\hat{j}_p^2v_p^2,
\label{33}
\end{eqnarray}

\noindent lead to results different from those of ref. \cite{Krm96}

\begin{eqnarray}
N=\langle \tilde{0}|\hat{N}|\tilde{0}\rangle&=&
\sum_n\hat{j}_n^2[v_n^2+(u_n^2-v_n^2){\cal N}_n], \nonumber\\
Z=\langle \tilde{0}|\hat{Z}|\tilde{0}\rangle&=&
\sum_p\hat{j}_p^2[v_p^2+(u_p^2-v_p^2){\cal N}_p].
\label{34}
\end{eqnarray}

\noindent In ordinary QRPA, where $|\tilde{0}\rangle \rightarrow |BCS\rangle$,
the states ${\cal O}^{11}_{\sss \pm}(JM)|BCS\rangle$ are null-vectors
and the F and GT sum rules are fulfilled, \ie the quantity
corresponding in this approximation to eq. (\ref{32}) vanishes.

\subsection {Self-consistency between the Residual Interaction and
the Mean Field}
\label{sec2.3}

It is well known that in the limit of exact isospin symmetry all the
$S_-(J^{\pi}=0^{\sss +})$ strength is concentrated in the isobaric
analog state (IAS), there is no $\beta^{\sss+}$ strength and the
$\beta\beta_{2\nu}$ decay is forbidden, \ie

\begin{equation}
S_{\sss IAS}\equiv S_{\sss -}(J^{\pi}=0^{\sss +}), ~~~~S_{\sss
+}(J^{\pi}=0^{\sss +})\equiv 0, ~~~~{\cal M}_{2\nu}(J^{\pi}=0^{\sss
+})\equiv 0.
\label{35}
\end{equation}

\noindent The question of which conditions have to be fulfilled in
order for these relations to hold in the QRPA has been discussed in
ref. \cite{Hir90a}. Within the RQRPA they read

\begin{eqnarray} 
{\sf X}_{pn}^{\sss IAS}&=&
D_{pn}^{1/2} \Lambda^{\sss +}_{pn}(J^{\pi}=0^{\sss +})/\sqrt{N-Z}=
\sqrt{N-Z} D_{pn}^{1/2} \hat{j_p}u_pv_n,
\nonumber \\
{\sf Y}_{pn}^{\sss IAS}&=&-D_{pn}^{1/2}
\Lambda^{\sss +}_{pn}(J^{\pi}=0^{\sss +})/\sqrt{N-Z}=
-\sqrt{N-Z} D_{pn}^{1/2} \hat{j_p}u_nv_p,
\label{36} \end{eqnarray}

\noindent as can be checked from eqs. (\ref{16}) and (\ref{18}). Putting
these expressions into the RQRPA equations \rf{2} we get

\begin{eqnarray} 
\epsilon_p+\epsilon_n -\omega_{\sss IAS}&=&
 -U_{j_p=j_n}^{\sss IAS}+ \frac{v_p}{u_p}
\Delta_{p}^{\sss IAS}+\frac{u_n}{v_n} \Delta_{n}^{\sss IAS},
\nonumber\\
\epsilon_p+\epsilon_n +\omega_{\sss IAS}&=&
U_{j_p=j_n}^{\sss IAS}+ \frac{u_p}{v_p}
\Delta_{p}^{\sss IAS}+\frac{v_n}{u_n} \Delta_{n}^{\sss IAS}.
\label{37} \end{eqnarray}

\noindent where

\begin{eqnarray} 
\Delta_{p}^{\sss IAS}&=&
-\frac{1}{2}\sum_{p'}\hat{j}_{p'}\hat{j}_{p}^{-1}
G(pp,p'p',J^{\pi}=0^{\sss +}) u_{p'}v_{p'} D_{ p'p'},
\nonumber\\
\Delta_{n}^{\sss IAS}&=&
-\frac{1}{2}\sum_{n'}\hat{j}_{n'}\hat{j}_{n}^{-1}
G(nn,n'n',J^{\pi}=0^{\sss +}) u_{n'}v_{n'} D_{ n'n'},
\label{38}\\
U_{j_p=j_n}^{\sss IAS}&=&\sum _{j_{p'}=j_{n'}}\hat{j}_{p'}\hat{j}_{p}^{-1}
F(pn,p'n',J^{\pi}=0^{\sss +}) (v^2_{n'} -v^2_{p'}) D_{p'n'}.
\nonumber
\end{eqnarray}

\noindent In deriving the eqs. \rf{37} the relation
$2G(pn,p'n',J^{\pi}=0^{\sss +}) =G(pp,p'p',J^{\pi}=0^{\sss +})
=G(nn,n'n',J^{\pi}=0^{\sss +})$ has been used. We have also assumed that
$J^{\pi}= 0^{\sss +}$ is the only important multipole affecting the
solution of eq. \rf{11}. This leads to the identity $D_{pn}=D_{pp}=D_{nn}$.

Summing and subtracting the two equations in \rf{37} we also obtain

\begin{eqnarray} 
\epsilon_{p} +\epsilon_{n}&=&
\frac{\Delta_{p}^{\sss IAS}}{2v_{p}u_{p}}
+\frac{\Delta_{n}^{\sss IAS}}{2v_{n}u_{n}},
\nonumber\\
\omega_{\sss IAS}&=& U_{j_p=j_n}^{\sss IAS}+
(e_p-\lambda_p)\frac{\Delta_{p}^{\sss IAS}}{\Delta_p}
-(e_n-\lambda_n)\frac{\Delta_{n}^{\sss IAS}}{\Delta_n}.
\label{39} \end{eqnarray}

\noindent As $\epsilon_{t}= \Delta_{t}/2v_{t}u_{t}$, the first relation
in \rf{39} will be an identity only if $\Delta_{t}={\Delta_{t}^{\sss IAS}}$.
This means that in solving the gap equations the substitution

\begin{equation}
G(tt,t't',J^{\pi}=0^{\sss +})\rightarrow
G(tt,t't',J^{\pi}=0^{\sss +}) D_{ t't'}
\label{40} \end{equation}

\noindent has to be done.

Also from the last relation in \rf{39} we obtain
\begin{equation}
\omega_{\sss IAS} +\lambda_p -\lambda_n =e_p-e_n + U_{j_p=j_n}^{\sss IAS}.
\label{41} \end{equation}

\noindent The left hand side in this equation is just the excitation energy
of the IAS relative to the ground state of the initial nucleus, \ie
$E_{\sss IAS}-E_i$.
\footnote{It should be remembered that in the BCS approximation the energy
difference between the ground states of an even-even $(N,Z)$ nucleus
and an odd-odd
$(N-1,Z+1)$
nucleus is $E_{\sss N-1,Z+1}- E_{\sss N,Z}=\Delta_p+\Delta_n+\lambda_p
-\lambda_n$; also $E_{\sss N-2,Z+2}- E_{\sss N,Z}=2(\lambda_p -\lambda_n)$.}
Further $e_p-e_n=\Delta_C-U_{j_p=j_n}^{\sss sym}$, where
$\Delta_C$ and $U_{j_p=j_n}^{\sss sym}$ are, respectively, the Coulomb
displacement energy and the symmetry energy. Thus, $E_{\sss IAS}-E_i$ will be
equal to $\Delta_C$ only if $U_{j_p=j_n}^{\sss sym} =U_{j_p=j_n}^{\sss IAS}$.
But, in the BCS approximation, the symmetry energy reads

\begin{eqnarray}
U_{j_p=j_n}^{\sss sym}&=&\sum _{j_{p'}=j_{n'}}\hat{j}_{p'}\hat{j}_{p}^{-1}
F(pn,p'n';0^{\sss +}) (v^2_{n'} -v^2_{p'})
\nonumber\\
&=&\frac{1}{2}\sum _{j'_p=j'_n} \sum_{JT}(-)^{T+1}\hat{J}^2\hat{j}^{-2}
G(jj'jj';JT) (v^2_{j'_n} -v^2_{j'_p}),
\label{42}\end{eqnarray}

\noindent and is equal to

\begin{equation}
U_{j_p=j_n}^{\sss sym}= \mu_{j_n} -\mu_{j_p},
\label{43} \end{equation}

\noindent where

\begin{eqnarray}
\mu_{j_n}&=& \sum_{n'}\hat{j}_n^{-1}\hat{j}_{n'}v_{n'}^2 F(nn,n'n';0^+)
+\sum_{p'}\hat{j}_n^{-1}\hat{j}_{p'}v_{p'}^2 F(nn,p'p';0^+)
\nonumber\\
&=& \frac{1}{2}\sum_{j'_p=j'_n} \sum_{J}\hat{J}^2 \hat{j}^{-2}
[G(jj'jj';JT=1)(2v_{j'_n}^2+v_{j'_p}^2)+G(jj'jj';JT=0)) v_{j'_p}^2 ],
\nonumber\\
\mu_{j_p}&=& \sum_{p'}\hat{j}_p^{-1}\hat{j}_{p'}v_{p'}^2 F(pp,p'p';0^+)
+\sum_{n'}\hat{j}_p^{-1}\hat{j}_{n'}v_{n'}^2 F(pp,n'n';0^+)
\nonumber\\
&=& \frac{1}{2}\sum_{j'_p=j'_n} \sum_{J}\hat{J}^2 \hat{j}^{-2}
[G(jj'jj';JT=1)(2v_{j'_p}^2+v_{j'_n}^2)+G(jj'jj';JT=0)) v_{j'_n}^2 ].
\label{44}\end{eqnarray}

\noindent are, respectively, the neutron and proton
self-energies.
This suggests that, when the self-energies are included in a
RQRPA calculation, the substitution

\begin{equation}
F(tt,t't',J^{\pi}=0^{\sss +})\rightarrow
F(tt,t't',J^{\pi}=0^{\sss +}) D_{ t't'}
\label{45} \end{equation}

\noindent is also pertinent.

\section {Results and Discussion}
\label{sec3}

The numerical calculations reported below are
performed with two-body delta-force and the single particle energies
used previously \cite{Hir90a}.  We define the ratios

\[
{\rm s}=\frac{{\it v}_s^{pp}}{{\it v}_s^{pair}}\quad;\quad
{\rm t}=\frac{{\it v}_t^{pp}}{{\it v}_s^{pair}},
\]

\noindent between the $T=1$, $S=0$ and $T=0$, $S=1$ coupling constants in the
{\it pp} channels and the pairing force constant, respectively. Their
physical values are ${\rm s}\cong 1.0$ and ${\rm t}\cong 1.5$. An
eleven dimensional model space, including all the single-particle
orbitals from oscillator shells $3\hw$ and $4\hw$ plus $0h_{9/2}$ and
$0h_{11/2}$ from the $5\hw$ oscillator shell, is used.

The results for two different RQRPA calculations will be discussed,
namely:

\begin{enumerate}
\item {\it Approximation I (AI)}: the BCS equations are solved in the
usual way with the constraint \rf{33}.
\item {\it Approximation II (AII)}: the pairing matrix elements are
renormalized as indicated in \rf{40} and the constraint \rf{34} is
used.
\end{enumerate}

\noindent As done in ref. \cite{Toi95}, we analyzed the difference
between the {\it double iteration} ({\it DI}) procedure, when the
proton and neutron occupations are given by \rf{11}, and the {\it
simple iteration (SI)} procedure, when the coefficients $D_{pn}$ on
the right hand side of eq.  \rf{11} are put equal to unity. Besides,
the influence of different number of multipolarities $J^{\pi}$ in the
summations involved in eqs. \rf{11} was studied by performing
calculations either with only the $J^{\pi}=1^+$ state (calculations
denoted {\it SI1} and {\it DI1}) or with the states
$J^{\pi}=1^+,2^-,3^+$ and $4^-$ (calculations denoted {\it SI4} and
{\it DI4}). The quantities that are of interest for the discussion
are: the amount of ground state correlations, defined as

\begin{equation}
C=\sum_p {\cal N}_p +\sum_n{\cal N}_n,
\label{46}
\end{equation}

\noindent the normalized GT strengths

\begin{equation}
s_{\sss -}= \tilde{S}_{\sss -}(J^{\pi}=1^+)/(N-Z);~~~~~
s_{\sss +}= \tilde{S}_{\sss +}(J^{\pi}=1^+)/(N-Z),
\label{47}
\end{equation}

\noindent the normalized GT sum rule,

\begin{equation}
{\sf S}_{\sss R}=s_{\sss -}-
s_{\sss +},
\label{48}
\end{equation}

\noindent the lowest energy eigenvalue $\omega_{J^{\pi}=1^+}$ and the
$\beta\beta_{2\nu}$ matrix element given by \rf{21}.

As an example, we first discuss the calculations of 
these observables in $^{76}Ge$ nucleus. They are shown in Fig. 1
for different {\it AI}'s ({\it SI1, DI1, SI4} and {\it DI4}), as a
function of the parameter {\rm t}. The QRPA results, up to the
collapse of this approximation, are also presented in the same
figure. One can immediately see that:

a) At variance with the QRPA, none of the RQRPA's collapses for
physically meaningful values of {\rm t}, and all four {\it AI}'s yield
quite similar results for $\omega_{J^{\pi}=1^+}$.

b) The difference between the {\it SI} and {\it DI} procedures is
always rather small.

c) The GSC generated by the states with multipolarities
$J^{\pi}\ne 1^+$ are not so significant as one would perhaps expect.

d) The violation of the Ikeda sum rule is more pronounced when more
GSC are taken into account. As the $\beta^+$-strengths within the
RQRPA and the QRPA are practically the same, the deviation from the
QRPA value ${\sf S}_{\sss R}=1$, mainly arises from the decrease of
the $\beta^-$-strength.

e) The different {\it AI}'s yield very similar results for the
$\beta\beta_{2\nu}$ matrix element ${\cal M}_{2\nu}(J^{\pi}=1^+)$, and
it is difficult to say which one is "better" and which one is "worse".
As a function of {\rm t}, they all seem to behave in the same way as
the corresponding QRPA matrix element. Still, the RQRPA results are
qualitatively different in the sense that they cannot be fitted by the
$(1,1)$-Pad\'e approximant

\begin{equation}
{\cal M}_{2\nu} \cong {\cal M}_{2\nu}({\rm t}=0)
\frac{1-{\rm t}/{\rm t}_0} {1-{\rm t}/{\rm t}_1},
\label{49}
\end{equation}

\noindent as happens in the QRPA case \cite{Krm93}.
(${\rm t}_0$ and ${\rm t}_1$ denote, respectively, the
zero and the pole of ${\cal M}_{2\nu}(J^{\pi}=1^+)$.) This is a direct
consequence of the absence of the collapse within the RQRPA for
the physical values of {\rm t}.

All the above comments are also valid for the nuclei $^{82}Se$,
$^{100}Mo$, $^{128}Te$ and $^{130}Te$. From now on we discuss the
differences between the {\it AI} and {\it AII} results. In doing this
we will recur only to the {\it SI} method. In Fig. 2 are compared the
behaviors of the Ikeda reduced sum rule ${\sf S}_{R}$ for all five
nuclei.
\footnote{It is worth noting that, when the {\it S1} procedure is used,
${\cal N}_n={\cal N}_p$ and therefore
$\tilde{S}_-(J^{\pi}=0^+)-\tilde{S}_+(J^{\pi}=0^+)=
\langle \tilde{0}| \hat N-\hat Z|\tilde{0}\rangle$.
Thus from \rf{32} we see that the F sum rule is
exactly conserved within {\rm AII}. Contrarily, in the {\rm AI} it is
broken by $\lsim 1\%$ for ${\rm s}=1$ and by $\gsim 10\%$ for ${\rm
s}=2$.  The F matrix elements ${\cal M}_{2\nu}(J^{\pi}=0^+)$ behave in
the same way as the GT matrix elements and pass through zero at ${\rm
s}\cong 1$.}

One immediately sees that in all the cases, except for $^{100}Mo$, the
violation of this sum rule is quite more pronounced in the {\it AI}
than in the {\it AII}.
\footnote{From the theoretical point of view, the ${\cal
M}_{2\nu}(J^{\pi}=1^+)$ matrix element in $^{100}Mo$ is in same sense
peculiar, because of the strong predominance of the
$[0g_{7/2}(n)0g_{9/2}(p);J^{\pi}=1^+]$ configuration.} Fig. 3 focuses
the crossings of the calculated ${\cal M}_{2\nu} (J^{\pi}=1^+)$ matrix
elements and the corresponding matrix elements deduced from the
experimental half-lives (including the experimental errors)
\cite{Krm94}. Again, as we do not know which value of the parameter
{\rm t} should be used in each case, it is hard to say that the RQRPA
is a better model than the QRPA. What is definitively clear is that the
sensitivity of ${\cal M}_{2\nu}(J^{\pi}=1^+)$ on this parameter is
unavoidable and that it is not a consequence of the collapse of the
QRPA.  It has been pointed out more than once \cite{Hir90,Krm94} that
the ${\cal M}_{2\nu}(J^{\pi}=1^+)$ amplitude goes to zero within the
QRPA because of the partial restoration of the Wigner spin-isospin
SU(4) symmetry. This is a result of two antagonistic effects: the
spin-orbit term in the mean-field potential that destroys the SU(4)
symmetry (since it singles out one spin direction over the other) and
the {\it pn} residual interaction that favors the $LS$ coupling over the $j-j$
coupling scheme \cite{Boh69}.  A physical criterion for fixing the
triplet {\it pp} coupling strength {\rm t}, based on the maximal
restoration of these symmetries, has also been suggested (${\it t}_{sym}$).
\footnote{By maximal restoration of the SU(4) symmetry we mean that
for ${\it t}={\it t}_{sym}$ the maximal concentration of the
$\beta^-$-strength within the GT resonance takes place and the
$\beta^+$-strength is minimal. In no way it connotes that the
$J^{\pi}=1^+$ states belong to a single SU(4) multiplet.}
In table 1 we compare the values of ${\rm t}_{sym}$, obtained using
the recipe just mentioned, with those that are necessary to reproduce
the experimental ${\cal M}_{2\nu}(J^{\pi} = 1^{+})$ amplitudes (${\rm
t}_{exp}$). The differences are of the order of 10\% what is quite
auspicious since the values of the nuclear parameter (both for
schematic and realistic interactions) are not known with such a
precision. We also remind that some additional degrees of freedom, not
considered in the present calculations, such as the quadrupole and
octupole charge-conserving vibrations can play an important role for
all the nuclei discussed here.  Besides, the contributions of
odd-parity nuclear operators, arising from the $p$-wave Coulomb
corrections to the electron wave functions and the recoil corrections
to the nuclear current, are also significant for the
$\beta\beta_{2\nu}$-decays of $^{128}Te$ and $^{130}Te$ nuclei \cite{Bar95}.

In summary, when the GSC are taken into account the collapse of the
QRPA does not develop in the physical region of the {\it pp}-strength
parameter \cite{Toi95,Sch96,Krm96}. Yet, the GSC only slightly
mitigate the strong dependence of the $\beta\beta_{2\nu}$ transition
amplitude on this parameter, which is set on by the partial
restoration of the SU(4) symmetry.  The price that is paid to avoid
the collapse within the QRPA is the non-conservation of the Ikeda sum
rule within the RQRPA.
This violation cannot be eluded and comes from the fact that the states
generated  by the action of the scattering part of the GT operator on
the RQRPA ground state is not contained in the model space.

\newpage
\begin{figure}
\begin{center} { \bf Figure Captions} \end{center}
\caption{ \protect \footnotesize
Behavior of several relevant quantities (defined in the text) within different
RQRPA-{\it AI} approximations in the case the $^{76}Ge$ nucleus, as a function
of {\rm t}. The QRPA results, up to the collapse of this approximation, are
also exhibited.
The matrix elements ${\cal M}_{2\nu}$ are given in units of $[MeV]^{-1}$.
\label{fig1}}
\caption{ \protect \footnotesize
Ikeda sum rule within the approximations I (upper panel) and II (lower panel),
when the simple iteration method is used with only one intermediate state.
\label{fig2}}
\caption{ \protect \footnotesize
Calculated matrix elements ${\cal M}_{2\nu}$
(in units of $[MeV]^{-1}$) for several nuclei, as a function of ${\rm t}$.
Dashed and dotted  curves correspond to the approximations I and II,
respectively, and the solid lines are the QRPA results.
The fill patterns indicate the experimental results.
\label{fig3}}
\end{figure}

\newpage

\begin{table}
\caption{Comparison between the values of the parameter {\rm t}
necessary to reproduce the experimental ${\cal M}_{2\nu}(J^{\pi}
= 1^{+})$ amplitudes (${\rm t}_{exp}$) and those that lead  to
maximal restoration of the SU(4) symmetry (${\rm t}_{sym}$).}
\begin{center}
\begin{tabular}{|c|ccccc|}
\hline
&{$^{76}Ge$}&{$^{82}Se$}&{$^{90}Mo$}&{$^{128}Te$}&{$^{130}Te$}\\
\cline{2-6}
QRPA&&&&\\ \cline{1-1}
${\rm t}_{exp}$    &1.37&1.35&1.52&1.32 &1.30 \\
${\rm t}_{sym}$    &1.23&1.27&1.47&1.40 &1.40 \\
\hline
{\rm AI}&&&&\\ \cline{1-1}
${\rm t}_{exp}$    &1.44&1.40&1.67&1.33 &1.32 \\
${\rm t}_{sym}$    &1.25&1.32&1.55&1.42 &1.42 \\
\hline
{\rm AII}&&&&\\ \cline{1-1}
${\rm t}_{exp}$    &1.42&1.37&1.65&1.32 &1.30 \\
${\rm t}_{sym}$    &1.25&1.30&1.55&1.40 &1.40 \\
\hline
\end{tabular}
\end{center}
\label{tab1}
\end{table}
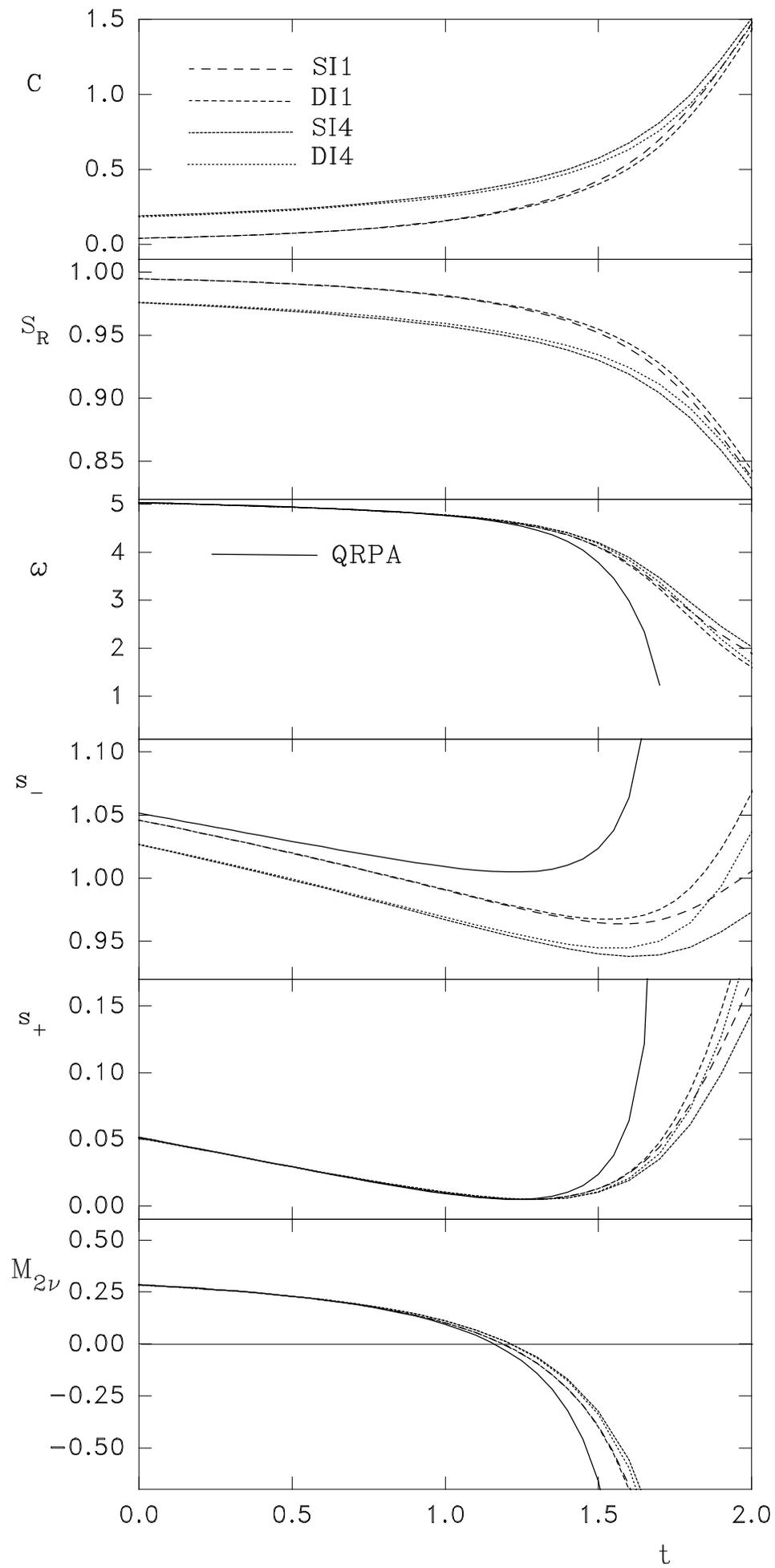
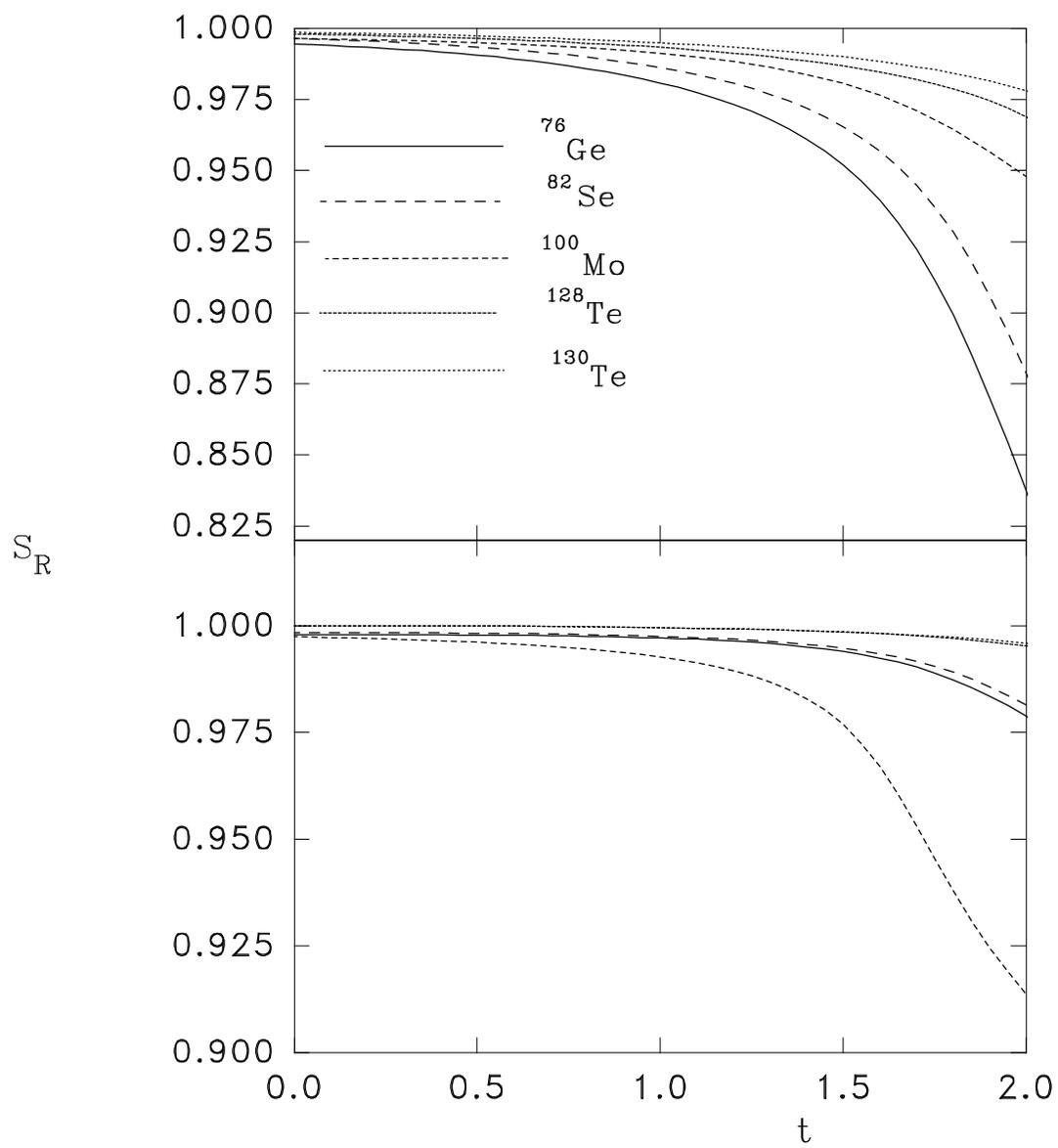
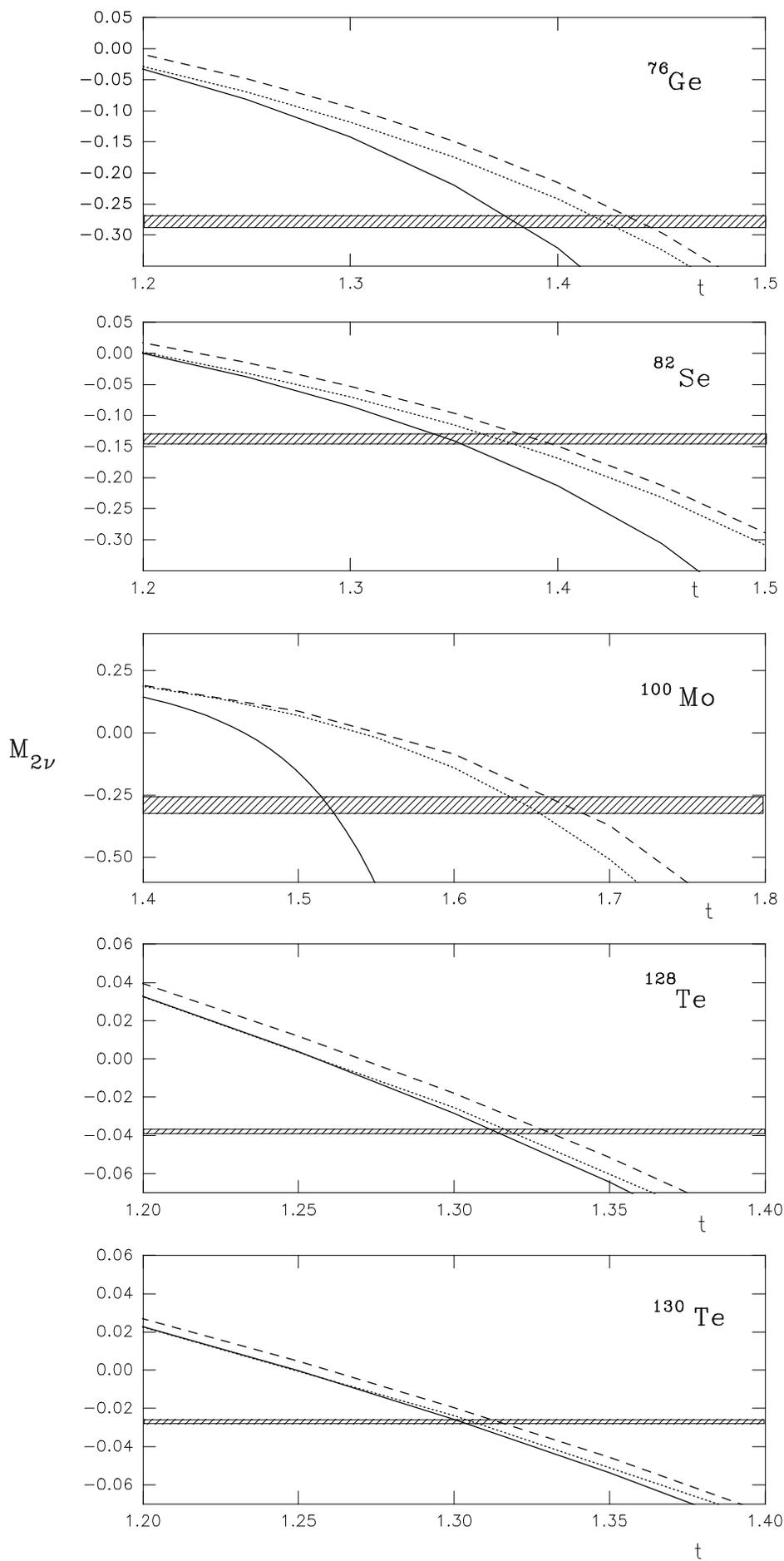
\end{document}